\documentclass[aps,prl,10pt,twocolumn,showpacs,preprintnumbers,amsmath,amssymb,superscriptaddress]{revtex4-1}
\usepackage[english]{babel}
\usepackage[dvips]{graphicx}
\usepackage{dcolumn}
\usepackage{bm}

\usepackage{color}

\begin{document}
\title{One-dimensional thermal pressure-driven expansion of a pair cloud into an electron-proton plasma}

\author{M. E. Dieckmann}
\affiliation{Department of Science and Technology (ITN), Link\"oping University, Campus Norrk\"oping, 60174 Norrk\"oping, Sweden}
\author{A. Alejo}
\affiliation{School of Mathematics and Physics, Queen's University Belfast, University Road, Belfast BT7 1NN, UK}
\author{G. Sarri}
\affiliation{School of Mathematics and Physics, Queen's University Belfast, University Road,
Belfast BT7 1NN, UK}
\author{D. Folini}
\affiliation{Universit\'e de Lyon, ENS de Lyon, CNRS, Centre de Recherche Astrophysique de Lyon UMR5574, F-69007, Lyon, France}
\author{R. Walder}
\affiliation{Universit\'e de Lyon, ENS de Lyon, CNRS, Centre de Recherche Astrophysique de Lyon UMR5574, F-69007, Lyon, France}

\email{Corresponding author: mark.e.dieckmann@liu.se}

\date{\today}
\begin{abstract}
Recently a filamentation instability was observed when a laser-generated pair cloud interacted with an ambient plasma. The magnetic field it drove was strong enough to magnetize and accelerate the ambient electrons. It is of interest to determine if and how pair cloud-driven instabilities can accelerate ions in the laboratory or in astrophysical plasma. For this purpose, the expansion of a localized pair cloud with the temperature 400 keV into a cooler ambient electron-proton plasma is studied by means of one-dimensional particle-in-cell (PIC) simulations. The cloud's expansion triggers the formation of electron phase space holes that accelerate some protons to MeV energies. Forthcoming lasers might provide the energy needed to create a cloud that can accelerate protons.  
\end{abstract}
\maketitle
Dense clouds of positrons and electrons were created in recent laser-plasma experiments \cite{Sarri15,Sarri17} and their evolution was examined. Such studies allow us to explore the physics of exotic plasmas found in the interior of imploding supermassive stars, close to accreting black holes and in the jets emitted by them \cite{Fender04,Smith07,Ruffini10,Marcowith16}. The x-ray precursor of the pair cloud created in Ref. \cite{Sarri17} ionized a background gas, which was contained in the experimental vessel prior to the laser shot ensuring the interaction with a pre-existing ambient plasma. The number of electron-positron pairs in this cloud was large enough to let the pairs behave as a plasma; a filamentation instability grew between the cloud and the ambient plasma \cite{Dieckmann10,Dieckmann15}. The limited size of the cloud implied that the electromagnetic field it carried could interact only during a short time with the ambient plasma, which ruled out any significant acceleration of the ions in the ambient plasma. 

Forthcoming experiments will generate larger and more energetic pair clouds that could potentially also accelerate ions. It is of interest to identify structures that unfold on electron time scales and can accelerate ions to large energies and to determine how large the pair clouds have to be in order to sustain these structures. 

Consider an electron-ion plasma with a small localized positive excess charge. The electrons oscillate in the positive potential of this space charge and a fraction of the electrons is trapped by it. Certain combinations of the distributions of the electric field, of the trapped electrons and of the untrapped electrons result in a stable solitary nonlinear wave known as electron phase space hole (EH) \cite{Bernstein57,Roberts67, Schamel79, Eliasson06}. It forms and survives on electron time scales for the plasma conditions found in laser-generated plasma \cite{Sarri10}. The electrostatic potential that sustains an EH might be capable of accelerating ions to high energies. 

Here we determine with particle-in-cell (PIC) simulations (See Ref. \cite{Arber15} for a description of the code) the energy, up to which protons can be accelerated by an EH, and the spatio-temporal scales over which the acceleration unfolds. We select initial conditions for the pair cloud and for the background plasma that are relevant for laser-plasma experiments. Restricting the simulation to one spatial dimension suppresses the growth of the competing filamentation instability, which simplifies our study of electrostatic instabilities. A restriction to one spatial dimension also stabilizes the EHs \cite{Newman01} and we can obtain an estimate for the maximum energy the protons can reach under idealized conditions. We perform several simulations, in which we vary the density ratio between the pair cloud and the ambient plasma in order to determine how it affects the proton acceleration.

The one-dimensional simulation box with the length $L_x$ and reflecting boundary conditions resolves the spatial interval $-0.3 \, L_x  \le x \le 0.7 \, L_x$. The pair cloud consists of electrons with the density $n_0$ and equally dense positrons and is located in the interval $x<0$. The number density $n_0$ yields the electron plasma frequency $\omega_p =  {(n_0e^2/m_e\epsilon_0)}^{1/2}$ ($e, m_e, \epsilon_0:$ elementary charge, electron mass and dielectric constant) that normalizes time ($t\rightarrow t\omega_p$) and the electron skin depth $\lambda_s = c/\omega_p$ ($c:$ speed of light) that normalizes space ($x\rightarrow x/\lambda_s$). The initial temperature of both species is $T_c = 400$ keV. The ambient plasma consists of electrons with the density $n_0/R$ and equally dense protons with the mass $m_p = 1836m_e$. The temperature of the electrons and protons is $T_a = 2$ keV. All electromagnetic fields are set to zero at the simulation's start. We examine the cases $R=8, 4, 1, 1/4$. We resolve $L_x=7520$ with $4 \times 10^4$ grid cells and each particle species by $N_p = 8 \cdot 10^7$ computational particles (CPs). The simulation evolves the plasma during the time $t_{sim}=3250$.

\begin{figure*}
\includegraphics[width=\textwidth]{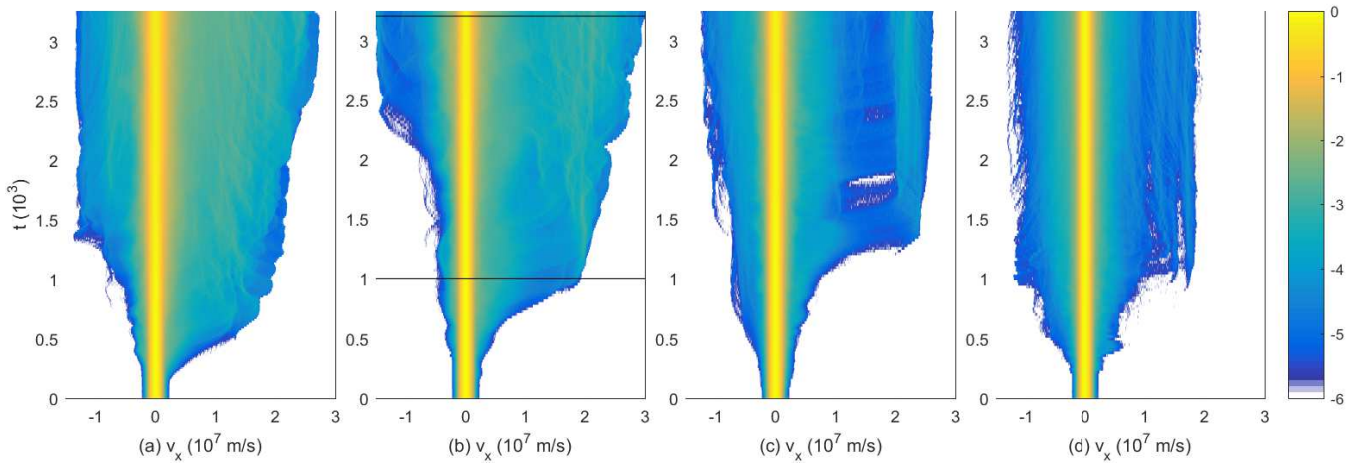}
\caption{Proton velocity distributions $\log_{10} f_p(v_x,t)$ normalized to their respective peak values: panel (a) shows the distribution for $R$=8, (b) that for $R$=4, (c) that for $R$=1 and (d) that for $R=0.25$. The color scale is clamped to -6. The horizontal lines in (b) indicate the times $t$ = 1000 and 3200.}\label{figure1}
\end{figure*}

Figure \ref{figure1} shows the time evolution of the proton velocity distributions for all four simulations. 
Protons are accelerated to MeV energies during a few $10^2$ time units in the case of the dense clouds and during about $10^3$ in the case of the dilute clouds. These proton energies exceed the thermal energy of the pair cloud by almost an order of magnitude. Protons are accelerated primarily along the expansion direction of the cloud. 

\begin{figure}
\includegraphics[width=\columnwidth]{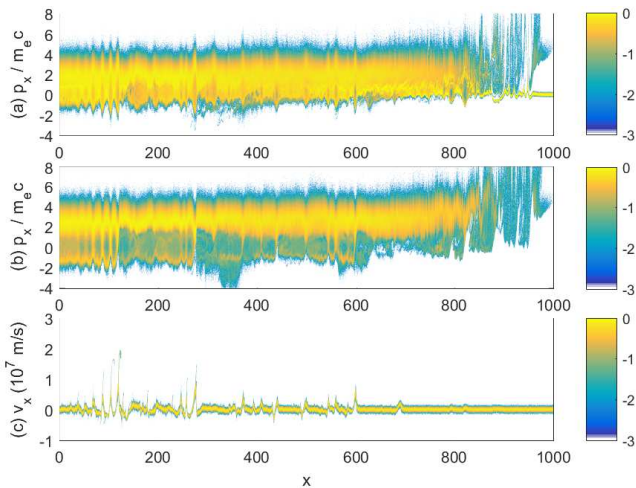}
\caption{Phase space density distributions for $R=4$ at $t=1000$: panels (a, b) show the distributions of the electrons and positrons normalized to the peak value in (a). Panel (c) shows the proton distribution normalized to its peak value. All color scales are 10-logarithmic (Multimedia view).}\label{figure2}
\end{figure}

We analyze the particle's phase space density distributions for the case $R=4$ in order to determine how the protons are accelerated. Figure \ref{figure2} shows the total electron distribution and those of the positrons and protons at the time $t=1000$.  
The cloud's front reached $x\approx 1000$. Its mean speed corresponds approximately to the cloud's initial thermal speed. The pairs with $x,v_x<0$ move to lower x and are reflected by the boundary.

We observe periodic trains of phase space holes in Figs. \ref{figure2}(a,b) at $40 < x < 140$, which coincide spatially with accelerating protons in Figure \ref{figure2}(c). The proton acceleration takes place at the boundaries of the EHs and the protons are accelerated primarily along increasing $x$. An EH corresponds to a localized positive excess charge, which is attractive for electrons and forms a barrier for positrons. The large EH at $x\approx 180$ in Fig. \ref{figure2}(a) blocks the low-energy positrons that flow to larger $x$ in Fig. \ref{figure2}(b) and their phase space density is strongly reduced in $150 < x < 230$. The reduced current contribution of the low-energy positrons in $150 < x < 230$ and the large current from the combined ambient and cloud electrons in this interval must be balanced by a faster motion of positrons; the positrons that overcome the EH potential are accelerated by it and the velocity distributions of electrons and positrons do not match in this interval. Similar processes are observed at $x\approx 270$ and $x\approx 600$.

The EHs far behind the cloud front form in a plasma that is composed of cool background electrons and of cloud electrons with a thermal spread that is larger than the relative speed between both populations. The electron acoustic instability \cite{Gary87} can form under such conditions. Linear instabilities tend to result in wave trains like the one in the interval $40<x<140$ in Fig. \ref{figure2}(a,b). We observe also solitary phase space holes in the electron distribution. These can either form when trains of phase space holes coalesce \cite{Roberts67} or in the presence of a net current, which could be caused by a local mismatch of the electron and positron currents \cite{Luque05}. 

The two-stream instability grows only if both electron species are cool \cite{Morse69} and move at a large relative speed. Such conditions exist at the cloud's front between the cloud electrons and those of the ambient plasma and this instability could be responsible for the train of phase space vortices in the interval $800 < x <10^3$. Figures \ref{figure2}(a,b) shows that the effect of the electrostatic two-stream instability, which grows at the cloud's front at $800 < x < 1000$, is to slow down the cloud electrons and to accelerate the positrons. On average the faster positrons will outrun the electrons at the cloud front, which results in a positive net current in this interval. This current can in the considered geometry only be balanced by a return current due to motion of the ambient electrons. Protons are hardly accelerated by the fast-moving phase space holes because they are exposed to their electric field only during a short time interval. 

\begin{figure*}
\includegraphics[width=\textwidth]{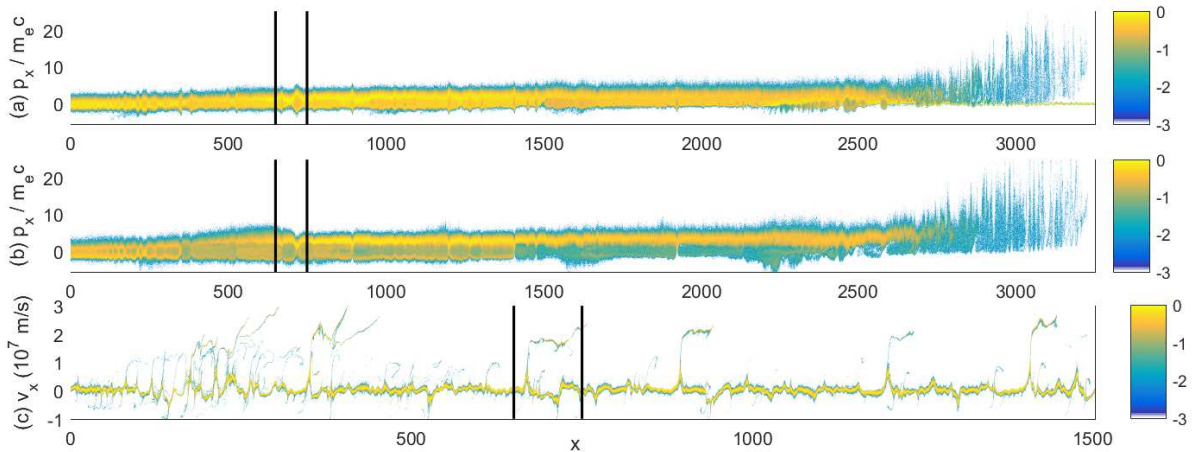}
\caption{Phase space density distributions for $R=4$ at $t=3200$: panels (a, b) show the distributions of the electrons and positrons normalized to the peak value in (a). Panel (c) shows the proton distribution normalized to its peak value. The displayed spatial interval in (c) is reduced compared to that in (a, b) in order to improve the resolution of the proton structures. All color scales are 10-logarithmic. The vertical lines denote $x$ = 650 and 750.}\label{figure3}
\end{figure*}

Figure \ref{figure3} shows the phase space density distributions at the time $t=3200$.
The fastest cloud particles have advanced to $x\approx 3200$ and some of them have reached a relativistic factor $\Gamma > 20$. They have been accelerated by their repeated interaction with the turbulent electric fields at the cloud's front. Positrons are faster than the electrons in the interval $x>2500$, which allows them to compensate the current of the denser combined distribution of the ambient and cloud electrons. Figure \ref{figure3}(a) reveals solitary EHs for all $x<2700$. The electric fields of the EHs at $x\approx 700$ and $x\approx 1400$ strongly modify the positron and proton distributions. Protons are accelerated to speeds up to $v_x \approx 2.5 \times 10^7$ m/s and these protons thus constitute the dilute population of energetic protons in Fig. \ref{figure1}(b). Protons are hardly accelerated for $x>2500$ (not shown) due to the relativistic flow speed of the electrostatic structures close to the cloud front.

\begin{figure}
\includegraphics[width=\columnwidth]{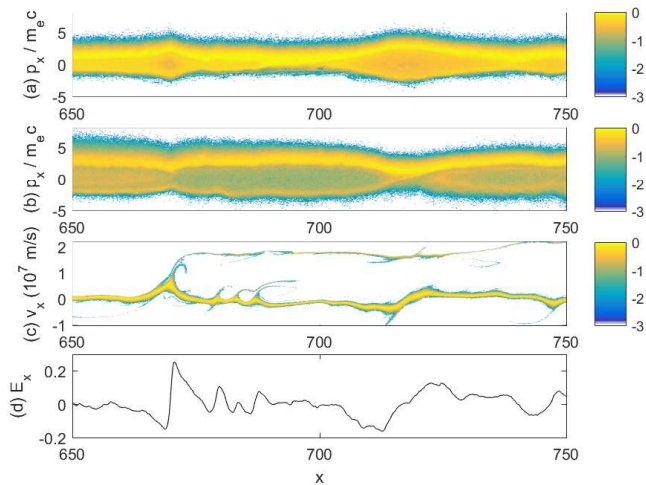}
\caption{Phase space density distributions and the electric field along the simulation direction for $R=4$ at $t=3200$: panels (a, b) show the distributions of the electrons and positrons normalized to the peak value in (a). Panel (c) shows the proton distribution normalized to its peak value. All color scales are 10-logarithmic. The electric field component along x in panel (d) is normalized to $m_e c \omega_p / e$ (Multimedia view).}\label{figure4}
\end{figure}

Figure \ref{figure4} shows the particle and electric field distribution in the interval $650<x<750$ at the time $t=3200$.
Two EHs are located in Fig. \ref{figure4}(a) at $x\approx 670$ and at $x\approx 720$. Both phase space vortices have a mildly relativistic velocity width. Their electric fields (See Fig. \ref{figure4}(d)) trap the electrons in a vortex. The bipolar electric field pulse at $x\approx 670$ is asymmetric. The negative electric field spike is sufficiently strong to reflect the low-energy positrons that flow from $x=650$ to larger values of $x$. The high-energy positrons are slowed down by it and they are re-accelerated when they leave the potential of the EH at $x\approx 675$. The phase space distribution of the electrons and positrons resembles that in Ref. \cite{Luque05}.

The positive electric field spike at $x\approx 670$ accelerates protons. Figure \ref{figure4} (Multimedia view) shows that the mean velocity of the EH is positive. The moving electric field is strong enough to reflect a small fraction of the ambient protons. The proton structure at $x\approx 670$ resembles an electrostatic shock \cite{Hershkowitz81}, which is here driven by the moving EH. The symmetric electric field distribution of the broader stationary EH at $x\approx 770$ modulates the proton distribution but its vanishing mean speed implies that it can not reflect protons.

The proton reflection by EHs is fast enough to explain the rapid formation of the dilute energetic proton population in Fig. \ref{figure1}. The peak energy the protons can reach by a single specular reflection is limited by the mean speed of the EH. The  speed of the reflected protons remains constant after the reflection (See Fig. \ref{figure4}(c)). The long-term stability of the EH implies that protons are continuously accelerated to this high energy. The number of accelerating protons remains constant but the number of accelerated protons increases in time, which explains the energetic band at $v_x \approx 2 \times 10^7$ m/s in Fig. \ref{figure1}(c). 

The mean speed of an EH in the rest frame of the electrons depends on its shape and size and it can vary within a limited velocity range close to the cloud's mean speed. The preferential speed of EHs in the simulation frame thus depends on the mean velocity of the electrons. The mean speed of the combined electron distributions of the cloud and of the ambient plasma is positive, which implies that on average EHs will propagate to increasing values of $x$. Reflection by EHs that move towards increasing $x$ explains the asymmetric proton distributions in Fig. \ref{figure1}. The mean speed value of the electrons and, hence, of the EHs is largest if the ambient plasma is dilute, which explains why protons are accelerated to the largest speeds in the simulations with $R$ = 4 and 8.

The simulation shows that the size of the cloud along its propagation direction has to be larger than 100 $\lambda_s$ for proton acceleration and for the selected initial conditions. We compare this estimate to the cloud size that might be within reach for forthcoming lasers. 

Pairs are created in the experiment by letting a primary electron beam collide with a target. The beam with the energy 600 MeV and charge 0.4 nC, which was produced by the Gemini laser with its energy $\sim$ 14 J in Ref. \cite{Sarri15}, generated $10^9$ pairs with sub-MeV energies and the number of pairs scaled almost linearly with the charge of the primary beam. The lasers at the Extreme Light Infrastructure (ELI-NP) will be able to deposit 200 J of energy in 20 fs, which can create a primary electron beam with the charge 2 nC and energy $\sim$ 20 GeV. The larger energy per electron of this beam triggers a more efficient pair production cascade, which could potentially generate $10^{11}$ pairs. Pairs are generated during about $10^{-13}$ s \cite{Sarri15,Sarri17} and propagate at a speed close to $c$. The length of the cloud is thus 30 $\mu$m. Let us assume that its lateral size is comparable and that the $10^{11}$ pairs are distributed uniformly across the box with the side length $30 \mu$m. The resulting $n_0 \approx 4 \times 10^{18}\mathrm{cm}^{-3}$ gives the electron skin depth $\lambda_s \approx 2.7 \mu$m. The side length of the box is 12 $\lambda_s$. 

It might be possible to observe the acceleration of protons on such an experiment and for the initial conditions we used in the simulation if the lateral extent of the pair cloud could be reduced to about 1-2 $\lambda_s$ while keeping the EH stable and the cloud charge neutral. Pair clouds that are large enough to match the length criterion exist in energetic astrophysical environments \cite{Ruffini10}.

In conclusion we have examined the expansion of a hot pair cloud into a cooler ambient plasma, which consisted of electrons and protons. The electrons of the cloud and of the ambient medium mixed via EHs. The presence of slow-moving protons implied that on average the positrons must flow faster than the electrons in order to enforce charge- and current-neutrality. The electrostatic potential of the EHs adjusted the currents. The low speed of EHs implied that their electric field was practically stationary in the rest frame of the protons. Some slow-moving EHs reflected protons and accelerated them to energies that exceeded the initial thermal energy of the pair cloud by the factor 5. 

More realistic 2D simulation studies have to test how efficiently protons can be accelerated if electrostatic and magnetic instabilities compete and if the EHs become unstable \cite{Newman01}. A 2D simulation, which resolves y, could follow the initial growth phase of the filamentation instability and reveal how the filament formation and the associated magnetic fields affect the EHs.

Acknowledgements: M. E. D. acknowledges financial support by a visiting fellowship of CRAL (Centre de Recherche Astrophysique de Lyon, CRAL, Universit\'e de Lyon). DF and RW acknowledge support from the French National Program of 
 High Energy (PNHE). GS wishes to acknowledge support from EPSRC (grant No: EP/N027175/1). The simulations were performed on resources provided by the Swedish National Infrastructure for Computing (SNIC) at the HPC2N computer centre.

\end{document}